\documentclass[prl,twocolumn,showpacs,preprintnumbers,amsmath,amssymb]{revtex4}

\usepackage{graphicx}
\usepackage{dcolumn}
\usepackage{bm}

\begin{document}
\addtolength{\topmargin}{2cm}

\title{High-order Dy multipole motifs observed in DyB$_2$C$_2$ with
resonant soft x-ray Bragg diffraction}

\author{A.M. Mulders$^1$, U. Staub$^1$, V. Scagnoli$^1$, S.W. Lovesey$^{2,5}$, E. Balcar$^3$, T. Nakamura$^4$,
A. Kikkawa$^5$, G. van der Laan$^6$ and J.M. Tonnerre$^7$}
 \affiliation{$^1$Paul Scherrer Institut, CH 5232 Villigen PSI,
Switzerland}
%\author{S.W. Lovesey$^2$}
\affiliation{$^2$ISIS, Rutherford Appleton Laboratory, Chilton, 
Oxfordshire OX11 0QX, United Kingdom}
%\author{E. Balcar$^3$}
\affiliation{$^3$Vienna University of Technology, Atominstitut, Stadionallee 2. A1020, Vienna, Austria}
%\author{T. Nakamura$^4$}%
\affiliation{$^4$SPring-8/JASRI, Mikazuki, Sayo, Hyogo 679-5198, Japan}%
%\author{A. Kikkawa$^5$}
\affiliation{$^5$RIKEN Harima Institute, Mikazuki, Sayo, Hyogo 679-5148, Japan}
%\author{G. van der Laan$^6$}
\affiliation{$^6$Daresbury Laboratory, Warrington, Cheshire WA4 4AD, UK}
%\author{J.M. Tonnerre$^7$}
\affiliation{$^7$CNRS Grenoble, 38042 Grenoble Cedex 9, France}

\date{\today}

\begin{abstract}

Resonant soft x-ray Bragg diffraction at the Dy $M_{4,5}$ edges has been exploited to study Dy multipole motifs in
DyB$_2$C$_2$. Our results are explained introducing the intra-atomic quadrupolar interaction 
between the core $3d$ and valence $4f$ shell. This allows us to determine for the first time higher order multipole 
moments of dysprosium $4f$ electrons and to draw their precise charge density. 
The Dy hexadecapole and hexacontatetrapole moment have been estimated at $-$20\% and +30\% of the quadrupolar moment, respectively. 
No evidence for the lock-in of the orbitals at $T_N$ has been observed, in contrast to earlier suggestions.
The multipolar interaction and the structural transition cooperate along $c$ but they compete in the basal plane
explaining the canted structure along [110].
 
\end{abstract}


\maketitle

Resonant x-ray Bragg scattering (RXS) has recently become a powerful tool in modern solid state physics
to investigate magnetic, orbital and charge order phenomena associated with electronic degrees of freedom. 
Especially in transition metal oxides, and, possibly, actinide compounds, various phenomena can occur 
simultaneously and it is not easy to identify the exact role and interplay of the various order parameters. 
However, in rare earth based compounds, orbital and magnetic order occur more autonomously from the lattice as the
$4f$ electrons are shielded by the outer $5d$ electrons. Electronic orbital motifs are usually labeled 
antiferroquadrupolar (AFQ) or ferroquadrupolar (FQ) but higher order multipoles are not established by observations.
Argon atoms in an excited state with $l$=3 show a significant presence of the
higher order multipole moments as concluded from electron scattering \cite{ar1}, but in the solid state multipoles beyond 
rank 4 have never been detected. Yet the alignment of higher multipoles to the surrounding charge in the lattice is likely.
For $4f$ electrons ($l$=3) multipoles up to rank 6 are present and this results in a 
undulated and aspheric charge density.
Which multipoles dominate these orbital order transitions remains a rather controversial topic 
as exemplified in CeB$_6$ \cite{plakhty_prb_2005}, NpO$_2$ \cite{kubo_prb_2005} and
URu$_2$Si$_2$ \cite{kiss_prb_2005}. 
Using soft x-ray resonant Bragg scattering (SXRS), we have obtained the first direct 
evidence of high-order Dy multipole moment motifs in DyB$_2$C$_2$. 

DyB$_2$C$_2$ has attracted much attention lately as its high AFQ ordering temperature $T_Q$=24.7~K 
allows to study this phenomenon conveniently.
At room temperature, DyB$_2$C$_2$ crystallizes in the tetragonal $P4/mbm$ structure  
and undergoes a structural transition with small alternating shifts 
of pairs of B and C atoms along $c$ at $T_Q$ \cite{adachi_prl_2002} which 
reduces the symmetry to $P4_2/mnm$ \cite{tanaka_jpcm_1999}.
Below $T_N$=15.3~K antiferromagnetic order (AFM) is observed
with complex moment orientations due to the underlying
orbital interaction \cite{yamauchi_jpsj_1999}.
RXS at the Dy $L_3$ edge has been used to elucidate Dy dipole and 
quadrupole motifs \cite{tanaka_jpcm_1999,hirota_prl_2000}. A dipole transition (E1) occurs 
between Dy $2p$ and $5d$ shells and a quadrupole transition (E2) between Dy $2p$ and $4f$ shells. 
The first and dominant process probes the quadrupolar order of the $5d$ states 
and the latter transition probes the quadrupolar order of the $4f$ states. 
The quadrupolar origin of the reflection has been confirmed by azimuthal scans (rotation around the Bragg wave vector).

\begin{figure}[!t]
\vspace{0cm}
\includegraphics[width=0.45\textwidth,angle=270]{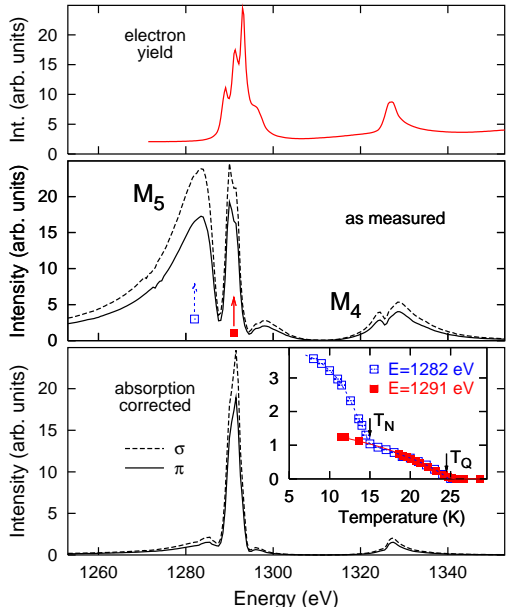}% Here is how to import EPS art
\vspace{-.3cm}
\caption{\label{fig_hv} Diffraction intensity of orbital ordering (00$\frac{1}{2}$) reflection in DyB$_2$C$_2$
taken with incident $\sigma$ and $\pi$ polarization at $T$=18~K (middle).
Top graph shows the absorption recorded in electron yield and the bottom graph 
shows the diffracted intensity after correction for absorption.
The inset shows the intensity as a function of temperature for $E$=1282~eV and 1291~eV.}
\label{fig_e}
\vspace{-.5cm}
\end{figure}

Ground state wave functions of the Dy $4f$ shell have been proposed based on the observed absence of 
E2 intensity at the (00$\frac{5}{2}$) charge forbidden reflection which implied cancellation of quadrupole (rank 2)
and hexadecapole (rank 4) contributions \cite{tanaka_prb_2004}.
These wave-functions are consistent with a recent inelastic neutron 
scattering study where the orbital fluctuation timescale has been determined \cite{staub_prl_2005}. 
An alternative explanation for the 
RXS data suggest that the hexadecapole moment is zero and the overlap between the E1 and
E2 resonances cancels the E2 intensity \cite{matsumura_prb_2005}. 
However, this analysis is not sound \cite{remark1}.
SXRS at the Dy $M_{4,5}$ edges accesses the 4f shell directly with the E1 
transition, {\it without} the complication of overlapping E1 and E2 resonances.
Our results show that the Coulomb (intra-atomic quadrupole) interaction
between the $3d$ and $4f$ shells is significant.
The ordered quadrupole moment of the $4f$ shell in the intermediate state reshapes the observed 
$3d_{5/2}$ and $3d_{3/2}$ electron density and leads to a core hole splitting.
It is shown that this phenomenon causes interference between different pathways of the
scattering  amplitude allowing the extraction of high-order multipole moments of the $4f$ shell
that have rank 4 and rank 6 (hexacontatetrapole).

A DyB$_2$C$_2$ single crystal has been grown by Czochralski
method using an arc-furnace with four electrodes and cut with (001) perpendicular to the sample surface. 
Subsequently, it was polished and aligned with 26$^\circ$ azimuthal angle. Zero degrees 
azimuth corresponds to alignment of the $b$ axis in the scattering plane. 
The orbital ordering (00$\frac{1}{2}$) reflection has been recorded at 
the Dy $M_{4,5}$ edges of DyB$_2$C$_2$ at the RESOXS end-station of the SIM beam-line at the
Swiss Light Source. The Dy $M_{4,5}$ absorption edges have been characterized with fluorescence yield (FY) and 
electron yield (EY) at RESOXS and the BL25SU beam line at SPring-8, respectively.

The energy profile of the charge forbidden (00$\frac{1}{2}$) reflection, together with x-ray absorption (XAS) data (EY), 
is shown in Fig.~\ref{fig_e}. 
Multiple features are observed with much larger energy spread than the multiplet structure of the XAS data, in contrast 
with the single oscillator recorded at the Dy $L_3$ edge \cite{tanaka_prb_2004}.
A similar energy profile has been reported for the $M_5$ edge of Ho at the magnetic
(0,0,$\tau$) reflection but the central sharp feature (solid symbol in Fig.~\ref{fig_e}) is absent \cite{spencer_jpcm_2005}. 
The authors argue that the large photo-absorption causes this `gap' at the 
$M_5$ edge because strong absorption is inherent to soft x-ray scattering. 
To characterize the width of the scattering $\theta$/2$\theta$-scans have been recorded 
for the (00$\frac{1}{2}$) reflection at each energy. Comparison with 
FY and EY data shows part of the width is due to the limited penetration depth. 
Subsequently, the integrated intensity has been corrected for absorption as shown in the bottom panel of Fig.~\ref{fig_e}. 
Absorption effects hugely reduce the intensity of the central feature but cannot 
account for the multiple features in the energy dependent intensity. 
The different character of those structures becomes even more apparent from the temperature dependence 
of the scattered intensity. The AFM transition is witnessed at 15~K as a gradual 
increase in intensity at 1282~eV, however this increase is absent at 1291~eV
(see inset of Fig.~\ref{fig_e}). This cannot be caused by absorption effects on a single oscillator resonance.
It appears that the scattered intensity at 1291~eV is sensitive to the orbital order only and its
gradual increase below $T_N$ shows that there is no lock-in of the orbitals as proposed from neutron
diffraction  and symmetry analysis \cite{zaharko_prb_2004}.

The structure in the energy profile is similar for $\sigma$ and $\pi$ polarization of the incident x-rays
and independent of the sample temperature in the AFQ phase. Azimuthal angles of 56$^\circ$ and 86$^\circ$ give the same 
spectral shape but show different overall intensity as expected from the azimuthal angle dependence of a quadrupole.
This illustrates merely one atomic tensor is active in the scattering process.
We propose that this particular shape of the energy dependence of the (00$\frac{1}{2}$) reflection is caused by the splitting 
of the $3d$ core states which results in multiple interfering resonators and consequently adds structure to the energy profile. 
Previously a core hole splitting caused by the intra-atomic magnetic 
interaction between $5f$ and $3d$ was used to describe the unusual double Lorentzian energy profile
at the $M_4$ resonance in NpO$_2$ \cite{lovesey_jpcm_npo2}. 
There is no ordered magnetism in the AFQ phase and we introduce 
the intra-atomic {\it quadrupolar} interaction to partially lift the core hole degeneracy.
The energy splitting is determined by the strength of the intra-atomic interaction 
and we show that the relative amplitudes are determined by the $4f$ wave function. 
Thus, the spectral shape is constant in the AFQ phase. Note that the 
diffracted intensity is caused by the bulk order and subject to
critical fluctuations while the energy splitting
is due to the local orbital moment of the $4f$ shell and independent of its orientation.
We will first encapsulate the resonant diffraction theory specific to our problem and
introduce the intra-atomic quadrupole interaction to portray the observed energy dependence.

The scattered photon intensity $d\sigma/d\Omega$ is proportional to 
the squared modulus of the amplitude $f$ \cite{lovesey_review}
\begin{equation}
\label{eq_f}
f=\sum_{K,Q} (2K+1)^{\frac{1}{2}} X_{-Q}^K \sum_q D_{Qq}^K(\alpha,\beta,\gamma) F_q^K,
\end{equation}
\noindent 
which is described as a pro\-duct of spherical tensors with rank $K$.
$ F_q^K $ with $-K \leq q \leq K$ describes the electronic response of the sample. 
$X_{-Q}^K $ with $-K \leq Q \leq K$ describes the geometry of the ex\-pe\-ri\-men\-tal set-up and 
polarization direction of the incident and reflected x-rays. 
Lastly, $ D_{Qq}^K(\alpha,\beta,\gamma)$ rotates $ F_q^K $ onto the coordinates of the experimental 
reference frame used for $X_{-Q}^K $  with Euler angles $\alpha,\beta,\gamma$. The Dy site symmetry of $2/m$ gives 
$q$=$\pm 2$ and $K$=2 due to the absence of charge ($K$=0) and time-odd (magnetic) ($K$=1) 
order at (00$\frac{1}{2}$) in the AFQ phase.

$F_q^K$ describes the atomic resonant process which is commonly represented by a harmonic oscillator.
In our particular case $F_q^K $ is a sum of several oscillators at the E1 resonance created by the splitting 
of the core state. The amplitude $A_q^K (\bar{J},\bar{M})$ of each oscillator is labeled
by the total angular momentum $\bar{J}$=$\frac{3}{2}$,$\frac{5}{2}$ and magnetic quantum number $\bar{M}$
of the core hole.

\begin{equation}
\label{eq_F}
F_q^K \propto \sum_{\bar{J},\bar{M}} \frac{r_{\bar{J}} A_q^K (\bar{J},\bar{M})}{E-\Delta_{\bar{J}}-\epsilon(\bar{J},\bar{M}) 
+ i \Gamma_{\bar{J},\bar{M}}}.
\end{equation}
\noindent Here, $E$ is the photon energy, $\Delta_{\bar{J}}$ is the difference in energy between the 
degenerate $3d_{\bar{J}}$ shell and the $4f$ empty states and $\Gamma_{\bar{J},\bar{M}}$ the 
lifetime of the intermediate state. 
$\epsilon (\bar{J},\bar{M})$=$[3 \bar{M}^2 - \bar{J}(\bar{J}+1)] Q_{\bar{J}}$ is the energy shift of the $3d$ core levels due to the 
intra-atomic quadrupole interaction $Q_{\bar{J}}$, similar as in M\"{o}ssbauer spectroscopy.
$Q_{\bar{J}}$ is a product of the $3d$ quadrupole moment and the $f$-electron electric field gradient 
experienced by the $3d$ electrons with $Q_{\pm \frac{3}{2}}/Q_{\pm \frac{5}{2}}$=7/3.

\begin{figure}
\includegraphics[width=0.28\textwidth,angle=270]{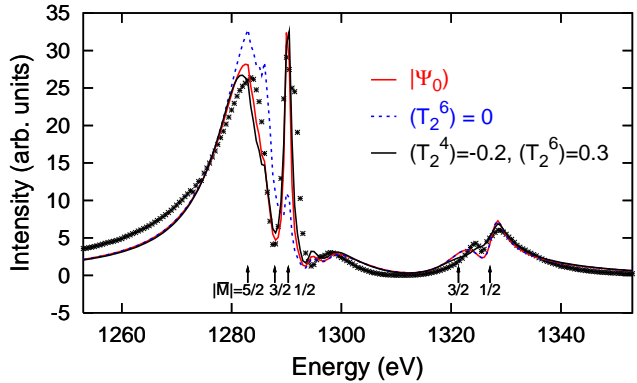}% Here is how to import EPS art
\caption{Observed energy profile (crosses) compared to fits of the theory including absorption correction with 
a) wave-function taken from Ref.~\cite{tanaka_prb_2004} (red), b) $\langle T_2^6 \rangle$ set to zero (dotted blue), 
and c) $\langle T_2^x \rangle$ as free parameter (black).
$\langle T_2^2 \rangle$ is normalized to 1, $\langle T_2^4 \rangle$=$-$0.2, $\langle T_2^6 \rangle$=+0.3,
$\Gamma_{\frac{5}{2},\pm \frac{1}{2}}$=0.8~eV,
$\Gamma_{\frac{5}{2},\pm \frac{3}{2}}$=2.7~eV,
$\Gamma_{\frac{5}{2},\pm \frac{5}{2}}$=5.3~eV,
$\Gamma_{\frac{3}{2},\pm \frac{1}{2}}$=1.9~eV and
$\Gamma_{\frac{3}{2},\pm \frac{3}{2}}$=5.4~eV.
}
\label{fig_fit}
\vspace{-.5cm}
\end{figure}

The amplitudes interfere and the branching ratio between the two edges is defined as a real mixing 
parameter $r_{\bar{J}}$.
$A_q^K (\bar{J},\bar{M})$ is constructed from the structure factor of the chemical unit cell $ \Psi_q^x$
\begin{eqnarray}
A_q^K (\bar{J},\bar{M})=(-1)^{\bar{J}-\bar{M}} \sum_r (2r+1) 
\left(
\begin{array}{ccc}
\bar{J} & r & \bar{J} \\
-\bar{M} & 0 & \bar{M}
\end{array}\right) \nonumber \\
\times  \sum_x
\left(
\begin{array}{ccc}
K & r & x \\
-q & 0 & q
\end{array}\right)
R^K (r,x) \Psi_q^x,
\label{eq_A}
\end{eqnarray}
\noindent where $R^K (r,x)$ are reduced matrix elements, $r$=0,1,...,2$\bar{J}$, $x$=$|K$$-$$r|, ...,|K$$+$$r|$ 
and $q$+$x$ and $r$+$x$ equal an even integer.
The Dy site symmetry, 2$/m$, dictates that $A_2^2$=$-A_{-2}^2$ and $
\Psi_q^x$=$\langle T_q^x \rangle - \langle T_{-q}^x \rangle$
where $\langle T_q^x \rangle$=$\langle \Psi_0 | T_q^x | \Psi_0 \rangle$, $T_q^x$ is the atomic spherical tensor of the Dy $4f$ shell 
and $| \Psi_0 \rangle$ its ground state wave function.
It follows that $A_q^K (\frac{5}{2},\bar{M})$ 
contains the hexadecapole moment $\langle T_2^4 \rangle$ and hexacontatetrapole moment 
$\langle T_2^6 \rangle$ in addition to the quadrupole moment $\langle T_2^2 \rangle$.
In case $\epsilon (\bar{J},\bar{M})$=0 the terms proportional to $\langle T_2^4 \rangle$ 
and $\langle T_2^6 \rangle$ cancel and $A_q^K (\bar{J})$ is proportional to $\langle T_2^2 \rangle$ as usual.

Let us here summarize the consequences of this analysis of the scattering. 
The overall intensity is well represented by scattering from the quadrupole 
$\langle T_2^2 \rangle$ of the 4f shell and reflects the same azimuthal angle dependence as 
the quantity observed in the dipole transition at the $L_3$ edge \cite{tanaka_prb_2004}. 
The dependence on the higher $4f$ multipoles arises from the splitting of the core hole states, which 
results in the $\bar{M}$ dependence of the amplitudes of the different harmonic oscillators (Eq.~\ref{eq_F}). 
Correspondingly, these amplitudes are influenced by higher $4f$ multipoles as shown in 
Eq.~\ref{eq_A} (Wigner-Eckart theorem). Therefore, the core hole splitting allows, for the first time, 
to measure the ordered hexacontatetrapole moment of an electronic shell in solid state physics.
Note, that this is not in contradiction with the fact that the integrated intensity of a 
dipole (E1) transition is sensitive to tensors up to rank 2 (quadrupole), while a quadrupole (E2) transition is sensitive to 
tensors up to rank 4 (hexadecapole).

Figure~\ref{fig_fit} shows the measured energy dependence compared to the above theory including absorption 
effect. This model describes the data well. $\Delta_{\bar{J}}$ typically depicts the onset of an
absorption edge and $r_{\frac{3}{2}}$ is calculated by Ref.~\cite{thole_prb_1985} at 0.22
which is in excellent agreement with our SXRS data. $Q_{\frac{5}{2}}$=$-$0.4~eV and the resonance positions are
indicated in Fig.~\ref{fig_fit}. The linewidth increases progressively from 0.8~eV for $\bar{M}$=$\pm \frac{1}{2}$ to 5.3~eV for
$\bar{M}$=$\pm \frac{5}{2}$. $|A_q^K (\frac{5}{2},\pm \frac{3}{2})|$ is relatively small.
Variations in $\langle T_2^x \rangle $ can be partially compensated by the line widths to yield a similar result. 
The best fit gives $\langle T_2^4 \rangle$=$-0.2$ and $\langle T_2^6 \rangle$=0.3 while
the $4f$ ground state wavefunction taken from Ref.~\cite{tanaka_prb_2004} gives 
$\langle T_2^4 \rangle$=$-$0.57 and $\langle T_2^6 \rangle$=0.36 and describes the data also well.
The corresponding fits as well as the effect of setting $\langle T_2^6 \rangle$ to zero are 
shown in Fig.~\ref{fig_fit} \cite{remark3}.
The temperature dependence observed at 1282~eV and 1291~eV supports our analysis since 
the resonant cross section contains additional time odd terms due to the magnetic order below $T_N$.
These terms may add at $\epsilon (\frac{5}{2},\pm \frac{5}{2})$ while they cancel
at $\epsilon (\frac{5}{2},\pm \frac{1}{2})$.

\begin{figure}
\includegraphics[width=0.46\textwidth,angle=0]{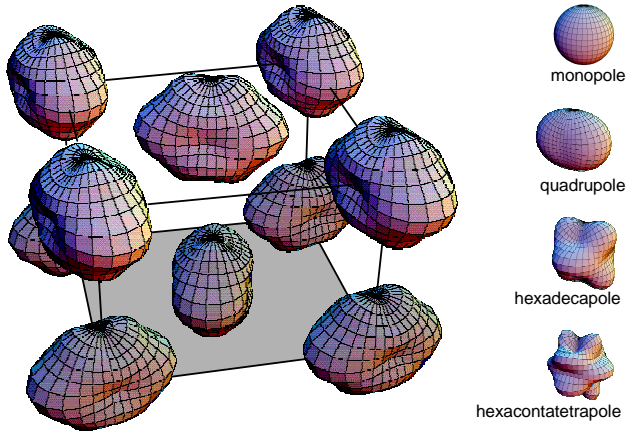}% Here is how to import EPS art
\caption{Dy charge density and multipole motif in the AFQ phase of DyB$_2$C$_2$. The basal plane is indicated in gray.
Note the 90$^\circ$ zig zag alignment of the Dy orbitals along $c$ and the canted zig zag 
alignment along [110].
A spherical charge density has been subtracted to emphasize the asphericity.
}
\label{fig_motif}
\vspace{-.5cm}
\end{figure}

The Dy $4f$ charge density obtained from the best fit is illustrated in Fig.~\ref{fig_motif}.
It supports a 90$^\circ$ zig zag alignment along $c$ as
demonstrated by its protrusions at top and bottom. 
In the basal plane a 90$^\circ$ zig zag alignment along [110] is preferred 
so that the concave part points toward the convex part of its nearest neighbor.
Yet, the absence of lock-in of the orbitals confirm that the orbital motif in the AFQ phase 
is similar to that in the AFQ+AFM phase and the Dy charge densities are canted away from [110] \cite{yamauchi_jpsj_1999}.
Such an AFQ structure logically emerges when the pairwise movement of B and C ions
is in competition with the multipolar interaction and promotes parallel alignment of the orbitals along [110] 
\cite{remark2}. In contrast,
the 90$^\circ$ zig zag alignment along $c$ is supported by both mechanisms. 
This is consistent with a recent study of (Dy,Y)B$_2$C$_2$  where quasi one 
dimensional AFQ along the $c$ axis has been reported \cite{indoh_jpsj_2004}.

We should add that our analysis in terms of core valence interaction 
is consistent with the multiplet structure interpretation.
In the M$_5$ and M$_4$ absorption the multiplet structure is $\sim$10 eV and $\sim$5 eV wide,
respectively, which is mainly due to the strong $3d$-$4f$ and $4f$-$4f$ Coulomb and exchange 
interactions \cite{thole_prb_1985}. The $3d^9_{5/2}4f^{10}$ and $3d^9_{3/2}4f^{10}$ final state multiplets 
contain 6006 and 4004 levels with an average lifetime width of 0.27 eV and 0.55 eV, respectively. 
The strong electron correlation leads to non-diagonal matrix elements in the
quantum numbers $\bar{L} \bar{S} \bar{J}$ of the core hole. 
The levels of the same $\bar{M}$ connected by these non-diagonal elements excite partly coherent. 
This makes it difficult to separate the core hole states.
In the free ion we do not expect an energy shift for distributions of different $|\bar{M}|$.
This was confirmed for the x-ray absorption spectrum with Cowan's atomic
Hartree-Fock code, where we found  roughly an equally broad multiplet when switching
off the quadrupole component of the orbital contribution of the core-valence interaction. This suggests that
any energy shifts in the $|\bar{M}|$ distributions will be due to multipolar ordering of the
$4f$ states.

In this theoretical framework the core hole state is uncoupled from the $4f^{n+1}$ state and
the effect of the $4f$ multiplet structure is indirectly taken into account by the effective 
widths of the $\bar{M}$ distributions, which exceed the intrinsic life time width and describes 
the data surprisingly well.

In conclusion, the Dy $4f$ hexadecapole and hexacontatetrapole moments in DyB$_2$C$_2$ have 
been measured with SXRS and their magnitudes are determined at $-$20\% and +30\% of the quadrupole 
moment, respectively. 
The orbital order remains unaffected by magnetic order below $T_N$
and no lock-in of the orbitals takes place, in contrast to CeB$_6$.
The structural transition and multipolar interactions cooperate along the $c$ axis while they compete along [110]. 
These findings amply demonstrate a new extension of the resonant x-ray Bragg diffraction
method for the observation of high-order electronic multipole motifs.
Of particular interest are 
materials in which higher order multipoles are believed to be of importance,  
such as CeB$_6$, URu$_2$Si$_2$ as well as the
$R$B$_2$C$_2$ and skutterudite families.

We thank O. Zaharko for valuable discussion. This work was supported by the Swiss National Science Foundation and 
performed at the SLS of the Paul Scherrer Institute, Villigen PSI, Switzerland.


\begin{thebibliography}{19}
\bibitem[{rem()}]{remark1}
\bibinfo{note}{The mixing parameter of E1 and E2 prosesses is fixed at unity in
  the analysis but this is not a priori justified.}
\end{thebibliography}
\end{document}